# Characterization and regulation of statistical properties in Er-doped random fiber laser

Xingyu Bao, Shengtao Lin, Jiaojiao Zhang, Yongxin Liang, Anchi Wan, Yifei Qi, and Zinan Wang*

Key Laboratory of Optical Fiber Sensing and Communications, University of Electronic Science and Technology of China, Chengdu 611731, China

*znwang@uestc.edu.cn

**Abstract:** Er-doped random fiber laser (ERFL) is a complex physical system, and understanding its intrinsic physical mechanisms is crucial for promoting applications. In this paper, we experimentally investigate the time-domain statistical properties of ERFL under full-bandwidth condition for the first time. We also analyze the effects of the transmission process and amplification process on the output characteristics of ERFL, on the basis of which we realize its regulation. This study guides RFL systems requiring transmission and amplification, offering fresh insights for regulating the time-domain stability.

## 1. Introduction

Random fiber laser (RFL) utilizes Rayleigh scattering along optical fiber to provide random distributed feedback with high efficiency, wavelength agility, high power output, et al. [1]–[4], which has attracted wide attention in the fields of laser inertial confinement fusion [5], optical fiber sensing [6], and imaging [7], also complex physical research [8], [9]. As an important branch of RFL, Er-doped random fiber laser (ERFL) utilizes Er-doped fiber (EDF) to provide gain [10]–[12], which is effective in increasing the output power and lowering the laser excitation threshold. In recent years, the research on ERFL has begun to expand to the deeper physical phenomena and mechanisms [13], [14], and it is worth to note that ERFL is also an excellent platform for the study of replica symmetry breaking, injection locking and other frontier problems [15]–[17]. In particular, the statistical properties have also attracted the attention of researchers [18], [19], which is extremely significant for the in-depth study of the physical mechanisms of random lasers. In previous studies, there are two approaches. The researchers have begun to investigate the statistical properties directly, but unable to capture high-frequency signals due to the limitation of bandwidths of the oscilloscope and photodetector used to obtain time-domain signals, which means the actual properties can't be reflected accurately. While the limitation is broken by means of filtering spectrum. It leads to lack of the understanding of ERFL's overall statistical properties which is an important issue. The ideal research setting is characterized by a full-bandwidth condition, where the optical bandwidth remains smaller than the electrical bandwidth without any filtering applied.

In addition, ERFL is subject to dispersion and nonlinear effects when transmitting in optical fibers [20]–[22] which would significantly affect the output characteristics of ERFL. Besides, the amplification process is necessary for stably transmitting ERFL signals to compensate the transmission loss [23], [24], which also affects output characteristics of ERFL. Revealing such effects is necessary because it is of great value in understanding the physical mechanisms and applying ERFL properly, but this issue has not been investigated before. Simultaneously, the amplified random fiber lasing system exhibits suppressed temporal dynamics attributed to its distinct open cavity structure [25]. This characteristic enables the output spectrum to remain unaltered and the linewidth to remain nearly constant throughout the entire power scaling range. Therefore, this system is suitable for conducting research on the impact of temporal characteristics in transmission and amplification under full-bandwidth condition.

In this paper, we experimentally investigate the full-bandwidth time-domain statistical properties of ERFL and the effects of transmission and amplification processes for the first time. A narrow-band ERFL is designed to provide an experimental basis for studying the time-domain characteristics of the ERFL under full-bandwidth condition, which shows inward deviation from the exponential distribution. After that, we take ERFL as an example revealing that both the transmission and amplification could affect the time-domain statistical properties. For transmission, dispersion compensation is proposed to improve the stability, while in amplification, the degree of which needs to be chosen reasonably. This work is instructive for RFL systems that require transmission and amplification.

## 2. Full-bandwidth statistical properties of ERFL

The experimental setup of ERFL is shown schematically in Figure 1. The 10-m-long erbium doped fiber (EDF) is pumped by a 1455 nm Raman fiber laser through a 1455 nm/1550 nm wavelength division multiplexer (WDM), which provides the gain for the 1550 nm lasing. The inclusion of an isolator (ISO) between the pump and WDM serves the purpose of avoiding spurious back-reflections. The highly reflective fiber Bragg grating (HR-FBG) with 1550 nm center wavelength and 0.1 nm 3 dB bandwidth connects to the 1550 nm port of WDM, acting as a point reflector to compromise a half-open cavity with the random distributed feedback in the 3-km-long single mode fiber (SMF). The generated ERFL outputs at the end of SMF, and is separated from the pump by an additional 1455 nm/1550 nm WDM. For spectrum measurements, we use an optical spectrum analyzer (OSA) with 0.01 nm resolution while the time-domain signals are detected by a 40 GHz photodetector (PD) and sampled by a 16 GHz oscilloscope (OSC).

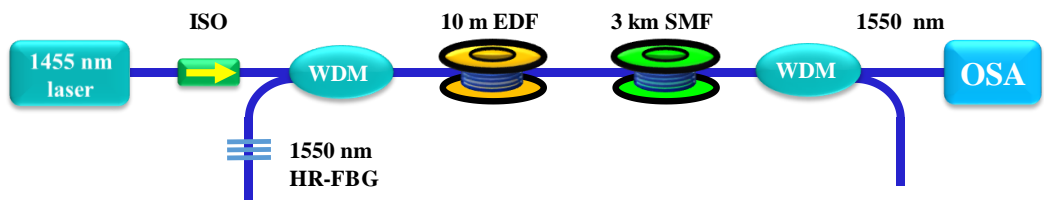

**Figure 1** Schematic of experimental setup for generating ERFL. ISO, isolator; HR-FBG, highly reflective fiber Bragg grating; WDM, wavelength division multiplexer; EDF, erbium doped fiber; SMF, single mode fiber; OSA, optical spectrum analyzer; PD, photodetector; OSC, oscilloscope.

We first characterize the power and spectral performance of ERFL, and the results are depicted in Figure 2. As shown in Figure 2(a), the optical signal-to-noise ratio (OSNR) is as high as ~50 dB, so the effect of the pump to the results can be ignored. Figure 2(b) shows the measured output power of the 1550 nm random lasing as a function of the launched pump power. We have identified five specific data points, highlighted in red, within the region of flat growth for this study. Notably, the maximum output power recorded is approximately 3 times greater than the minimum output power observed. The output spectra of selected points are presented in Figure 2(c). As the pump power increases from 1.0 W to 2.2 W, the spectrum broadens obviously at the pedestal (>20 dB) which is unlike the conventional RFL based on pure Raman gain mechanism, while there is only slightly broadening at the central lasing wavelength. To give a distinct comparison of the spectral linewidth evolution versus the launched pump power, the 3 dB, 10 dB and 20 dB bandwidth are calculated and given in Figure 2(d). The results indicate a significant widening of the 10 dB and 20 dB bandwidths, especially for the 20 dB bandwidth which increased substantially from ~0.24 nm (30.0 GHz) to ~1.17 nm (146.1 GHz), whereas the 3 dB bandwidth only increased from ~0.02 nm (2.5 GHz) to ~0.08 nm (9.9 GHz).

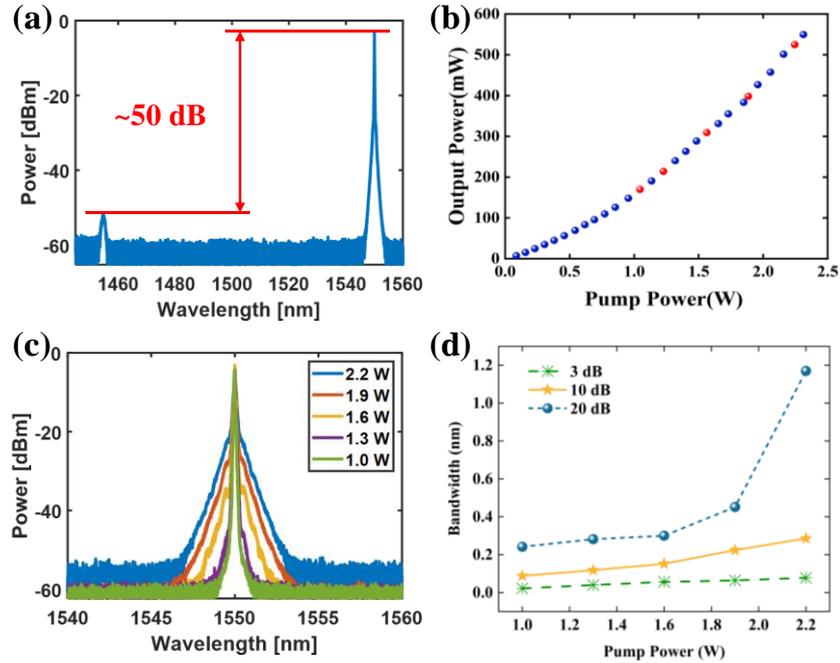

**Figure 2** Power and spectral properties of the ERFL. (a) Optical signal-to-noise ratio. (b) Output power versus pump power. (c) Output spectra at different pump powers. (d) Variation of 3 dB, 10 dB and 20 dB bandwidth of ERFL with pump power.

In order to measure the time-domain signals under full-bandwidth condition, the ERFL is recorded using a 40 GHz photodetector and a 16 GHz OSC, which are larger than the maximum 3 dB bandwidth (9.9 GHz) of the random lasing plotted in Figure 2(d). So the effect of frequency average does not exist, and we can obtain the real intensities and accurate statistics. For the purpose of analyzing the statistical properties, $2*10^8$ samples using the OSC at 20 GSamples/s sampling rate are acquired. We normalize the intensity I(t) to its mean value <I(t)> according to the experimental data. As shown in Figure 3(a), the ERFL exhibits minimal fluctuations on a sub-nanosecond timescale. The intensity probability density function (PDF) of selected points is illustrated on a vertical logarithmic scale in Figure 3(b). The black dashed line represents the exponential distribution, which is equal to the radiation consisting of statistically independent frequency components with Gaussian statistics. It can be seen that the intensity PDF deviates from the exponential distribution inwards indicting some correlations among different frequency components that exist in ERFL. Next, we shift our attention to the PDF's independence from variations in pump power, which remains the same as the pump power increases. Despite the output power rises to two times to its original value, the spectrum and 3 dB bandwidth only vary slightly within the detection bandwidth of the OSC (16 GHz) as shown in Figure 2(c) and 2(d). Consequently, the captured time-domain signals experience negligible changes, thereby preserving the nearly unchanged temporal statistical characteristics.

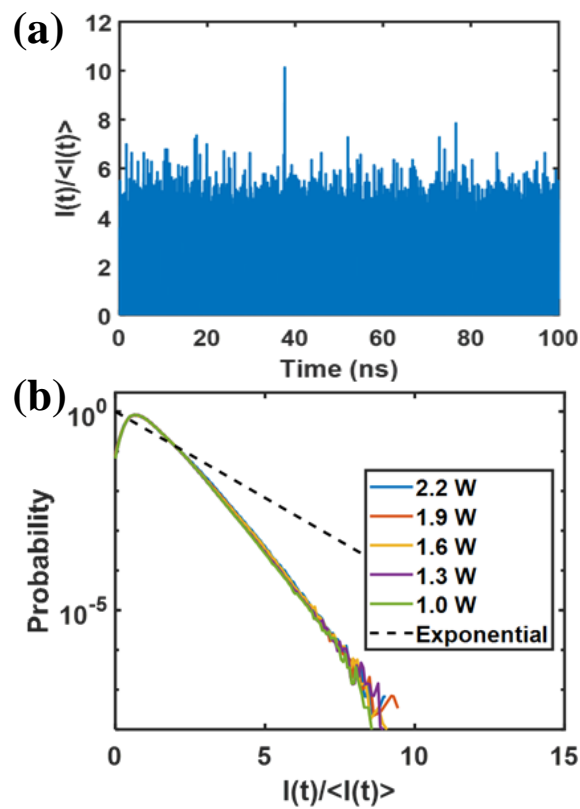

**Figure 3** (a) Temporal intensity dynamics. (b) Intensity PDF of the ERFL with pump power. The black dashed line represents the exponential distribution

## 3. Effects of transmission and amplification

The results above show that the generated ERFL has relatively good time-domain stability. Besides, ERFL exhibits the spectral broadening free property, which means the bandwidth of the amplified random fiber lasing remains almost the same with the seed source. Benefiting from these two properties, ERFL is a suitable basis to investigate the effects of transmission and amplification on the temporal statistical characteristics under full-bandwidth condition. The 3 dB bandwidth of the ERFL is 5 GHz when the pump power is 1.3 W. The ERFL at this case is selected as the seed source for the subsequent research.

## 3.1 Effects of the transmission

The experimental setup for investigating transmission's effects is presented in Figure 4. A section of SMF is connected to the ERFL. In the absence of any additional transmission components, the spectrum does not undergo broadening, hence detailed spectral representation is not provided here. After transmission, the signals are collected under full-bandwidth condition. The influence of transmission distance is considered and the length of SMF is taken as 25 km, 50 km and 75 km. Significant fluctuations are observed with the increase of transmission distance on sub-nanosecond time scale, as shown in Figure 5(a). After 25 km SMF, extreme events with intensity approximately 15 times higher than the average value can be recorded. Additionally, rare and intense events with peak powers up to 18 and 20 times the average power can be observed after traversing 50 km and 75 km of SMF, respectively. We use PDF to analyze statistical properties as shown in Figure 5(b). When compared to the ERFL seed source represented by the blue curve, the distribution curves of which after three lengths of SMF transmission all exhibit outward deviation. This variation suggests that the likelihood of extreme events increases after transmission, meaning that signals with intensity significantly larger than the mean value are more probable to occur. The fluctuations in intensity have increased, indicating a degradation in the time-domain stability. At the same time, it is important to note that, with the increase of the same transmission distance, i.e., 25km, the extent of outward deviation in the distribution curve is decreasing, indicating that there is an upper limit to the impact of transmission on the time-domain stability.

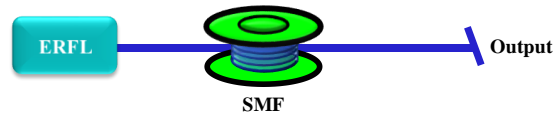

**Figure 4** Experimental setup for investigating transmission's effects.

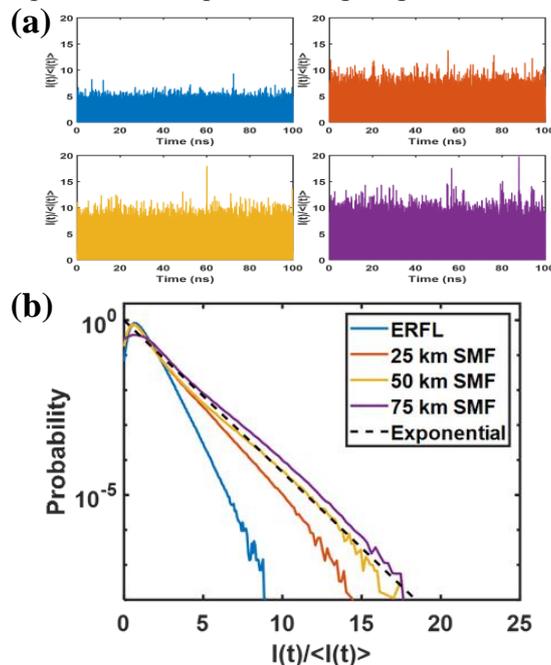

**Figure 5** (a) Temporal intensity dynamics and(b) Intensity PDF of the ERFL with different transmission distance.

This evolution may be related to the accumulation of fiber dispersion. To verify this idea, we performed experiments on dispersion compensation and the results of the PDF alteration are illustrated in Figure 6. Considering that the dispersion value of DCF is approximately 5 times that of SMF, we connected a 2.2 km dispersion compensating fiber (DCF) after the 25 km SMF and measured the signal under the same conditions. Same experiment is done on the ERFL seed source by connecting a section of 450 m DCF after it. The length of DCF is also selected based on the difference in dispersion value between it and SMF. Inside the ERFL seed source, there is a 3km-long SMF as depicted in Figure 1. The PDFs after dispersion compensation both veer inward as plotted in Figure 6(a) and 6(b), indicating that the time-domain stability has become better. The experimental results above indicate that the transmission process in the SMF deteriorates the time-domain stability of the ERFL, and there is an upper limit to the influence of the transmission distance. Choosing a suitable length of DCF for dispersion compensation based on the length of the SMF is an effective method to improve time-domain stability. This suggests that, according to the specific requirements of practical applications, it is possible to regulate the time-domain stability of the ERFL by increasing or suppressing the impact of the transmission.

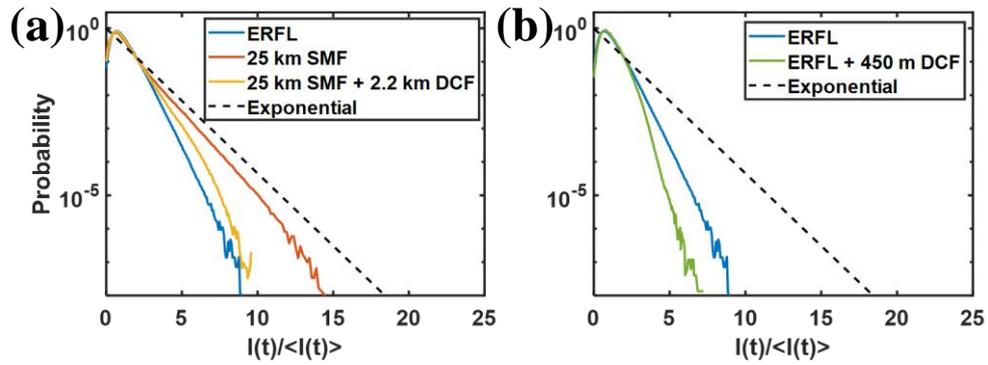

**Figure 6** Intensity PDF alteration under two cases of dispersion compensation.

## 3.2 Impacts of the amplification

Next, we study the impacts of amplification on the statistical properties and the experimental setup is presented in Figure 7. Similarly, generated ERFL with 3dB bandwidth of 5GHz is chosen as the seed source, which will be amplified by Master Oscillator Power Amplifier (MOPA) structure using another 1455 nm Raman pump. A 1550 nm highly reflective FBG combined with a circulator connected to the 50 km SMF for filtering out residual pump light. The 3dB bandwidth of the FBG is 1nm (~125 GHz), which significantly surpasses that of the ERFL seed source without introducing any detrimental effects to subsequent characterization analyses.

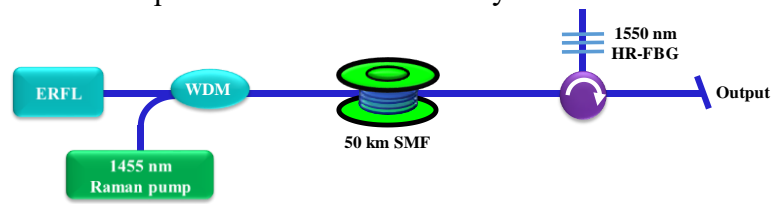

**Figure 7** Experimental setup for investigating amplification's effects.

The detailed lasing performance of the amplified ERFL output at 1550 nm is further analyzed, as illustrated in Figure 8. Figure 8(a) exhibits an OSNR exceeding 30 dB, which suggests that the influence of the pump can be disregarded. The output power versus the launched pump power is investigated, as shown in Figure8(b). We have selected four points for detailed analysis and marked them in red. We give slightly broadened spectra with increasing pump power in Figure 8(c), and it can be noticed that there is almost no broadening around the wavelength of 1550 nm as shown precisely in the inset of Figure 8(c). Figure 8(d) records the variation of the 3 dB bandwidth and 20 dB bandwidth for the corresponding spectra. The observed 3 dB bandwidths, which are all below 6 GHz as a result of the spectral-broadening-free property, consistently satisfy the full-bandwidth condition.

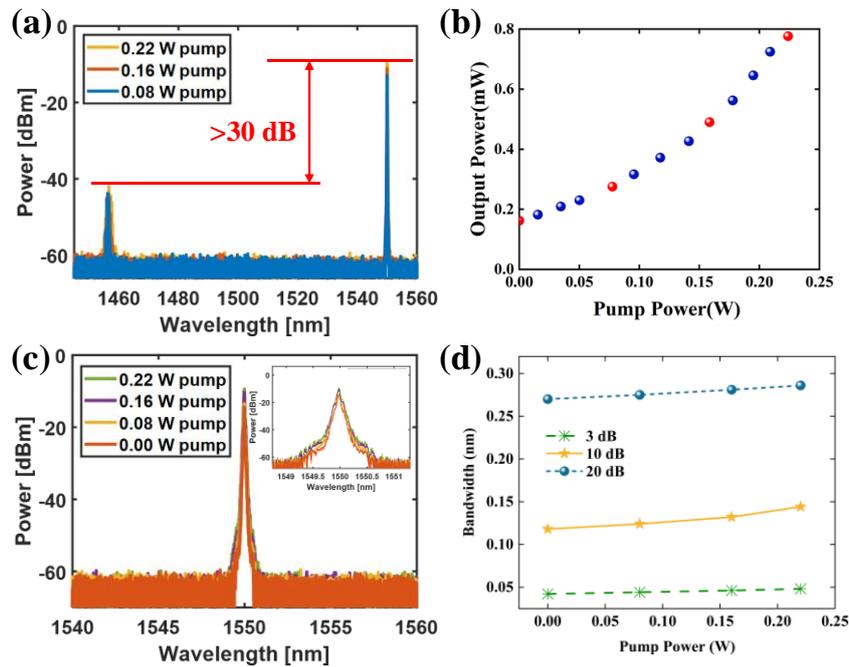

**Figure 8** Lasing performance of the amplified ERFL output. (a) Optical signal-to-noise ratio. (b) Output power versus pump power. (c) Output spectra at different pump powers. (d) Variation of 3 dB, 10 dB and 20 dB bandwidth of ERFL with pump power.

The time-domain signals are acquired at the output using the same photodetector and oscilloscope for analysis. Significant fluctuations are observed with the increase of the pump power on sub-nanosecond time scale, as shown in Figure 9(a). When the pump is 0.08 W, extreme events with intensity approximately 18 times higher than the average value can be recorded. Additionally, rare and intense events with peak powers up to 20 and 23 times the average power can be observed when the pump is 0.16 W and 0.22 W, respectively. We continue to analyze statistical properties using PDF as shown in Figure 9(b). As the degree of amplification increases, the slope of the distribution curve decreases indicating a signal which is much larger than the mean value is more likely to occur. That means, the ERFL's stability is getting worse after amplification.

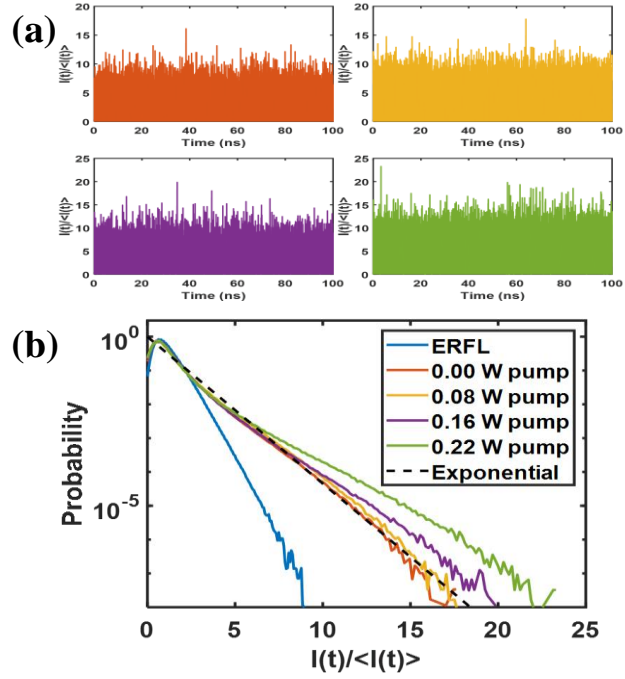

**Figure 9** (a)Temporal intensity dynamics and (b) Intensity PDF of the ERFL with different amplified pump power.

In order to validate the robustness of our findings, we replicated the experiments using additional lengths of SMF, specifically 5 km, 15 km, 25 km, and 75 km. Remarkably, consistent outcomes are obtained across all experimental configurations. This conclusion serves as a crucial guideline, highlighting the requirement for a balanced approach between amplification level and time-domain stability. It is imperative to carefully select a reasonable degree of amplification so that the unwanted effects like nonlinearity could be avoided.

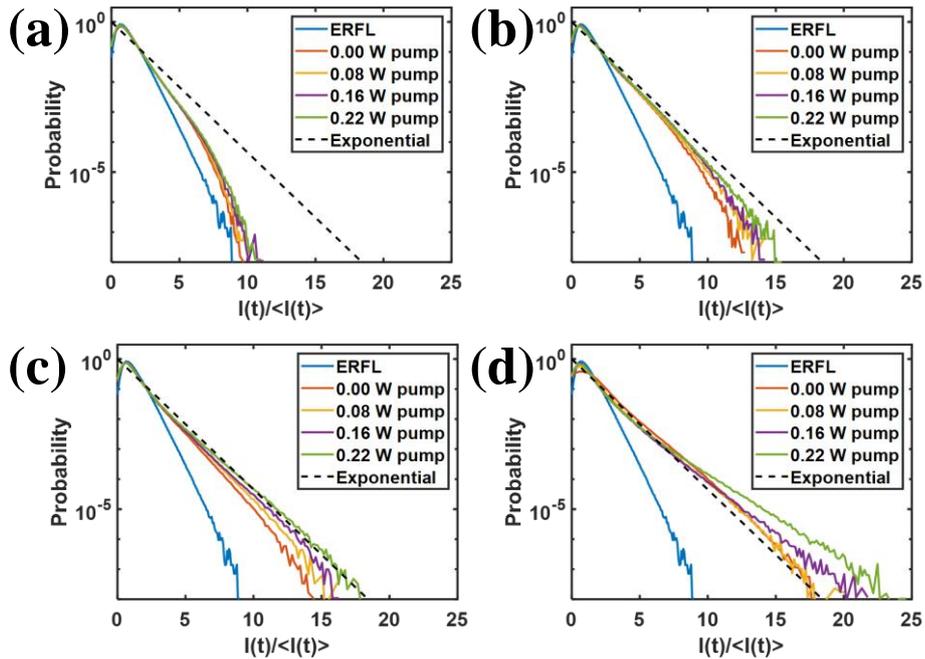

**Figure 10** Intensity PDF of the ERFL with different transmission distance.

This study examines the impact of transmission and amplification on time-domain statistical properties using ERFL as a case study. It is demonstrated that both transmission and amplification have notable effects on these properties. To enhance stability during transmission, dispersion compensation techniques are proposed. Additionally, in the context of amplification, selecting an appropriate degree is crucial for achieving desired performance. The findings presented in this work offer valuable guidance for RFL systems that require both transmission and amplification. Furthermore, this research contributes novel ideas for regulating time-domain stability.

## 4. Conclusions

In this paper, we present an experimental investigation of the time-domain statistical properties of ERFL under full-bandwidth conditions. Additionally, we examine the impact of the transmission process and amplification process on the output characteristics of ERFL. The experimental findings reveal a deviation from exponential distribution in the inward direction for the time-domain statistics of ERFL under the full-bandwidth condition. Subsequently, we utilize the ERFL as a case to demonstrate the influence of both transmission and amplification on time-domain statistical properties. For transmission, we propose dispersion compensation techniques to enhance stability, while in the context of amplification, it is crucial to select an appropriate degree for optimal performance. This study provides valuable guidance for RFL systems that necessitate both transmission and amplification. Besides, the present work also provides some new ideas for regulating the time-domain stability.

## References

[1] Z. Wang et al., "High Power Random Fiber Laser With Short Cavity Length: Theoretical and Experimental Investigations," IEEE J. Sel. Top. Quantum Electron., vol. 21, no. 1, pp. 10–15, Jan. 2015.

[2] V. Balaswamy, S. Ramachandran, and V. R. Supradeepa, "High-power, cascaded random Raman fiber laser with near complete conversion over wide wavelength and power tuning," Opt. Express, vol. 27, no. 7, p. 9725, Apr. 2019.

[3] L. Zhang, H. Jiang, X. Yang, W. Pan, S. Cui, and Y. Feng, "Nearly-octave wavelength tuning of a continuous wave fiber laser," Sci. Rep., vol. 7, no. 1, p. 42611, Mar. 2017.

[4] H. Zhang et al., "Quasi-kilowatt random fiber laser," Opt. Lett., vol. 44, no. 11, p. 2613, Jun. 2019.

[5] M. Fan et al., "Spectrum-tailored random fiber laser towards ICF laser facility," Matter Radiat. Extrem., vol. 8, no. 2, p. 025902, Mar. 2023.

[6] S. Lin, Z. Wang, Y. Qi, B. Han, H. Wu, and Y. Rao, "Wideband Remote-Sensing Based on Random Fiber Laser," J. Light. Technol., vol. 40, no. 9, pp. 3104–3110, May 2022.

[7] R. Ma, Y. Rao, W. Zhang, and B. Hu, "Multimode Random Fiber Laser for Speckle-Free Imaging," IEEE J. Sel. Top. Quantum Electron., vol. 25, no. 1, pp. 1–6, Jan. 2019.

[8] S. Lin et al., "Radiation build-up and dissipation in Raman random fiber laser," Sci. China Inf. Sci., 2023, doi: 10.1007/s11432-021-3329-6.

[9] J. Xu et al., "Optical rogue wave in random fiber laser," Photonics Res., vol. 8, no. 1, p. 1, Jan. 2020.

[10] Q. Meng, H. Wu, B. Han, J. Li, and Z. Wang, "LD-Pumped Random Fiber Laser Based on Erbium-Ytterbium Co-Doped Fiber," Photonic Sensors, vol. 10, no. 2, pp. 181–185, Jun. 2020.

[11] W. Zhang, Y. Song, X. Zeng, R. Ma, Z. Yang, and Y. Rao, "Temperature-controlled mode selection of Er-doped random fiber laser with disordered Bragg gratings," Photonics Res., vol. 4, no. 3, p. 102, Jun. 2016.

[12] N. Lizárraga, N. P. Puente, E. I. Chaikina, T. A. Leskova, and E. R. Méndez, "Single-mode Er-doped fiber random laser with distributed Bragg grating feedback," Opt. Express, vol. 17, no. 2, p. 395, Jan. 2009.

[13] B. C. Lima et al., "Observation of Lévy statistics in one-dimensional erbium-based random fiber laser," J. Opt. Soc. Am. B, vol. 34, no. 2, p. 293, 2017.

[14] B. Saxena, X. Bao, and L. Chen, "Suppression of thermal frequency noise in erbium-doped fiber random lasers," Opt. Lett., vol. 39, no. 4, p. 1038, Feb. 2014.

[15] A. S. L. Gomes et al., "Glassy behavior in a one-dimensional continuous-wave erbium-doped random fiber laser," Phys. Rev. A, vol. 94, no. 1, p. 011801, Jul. 2016.

[16] C. Chen, H. Wang, P. Lu, S. Mihailov, L. Chen, and X. Bao, "Self-injection locking of a low-noise erbium-doped random fiber laser by a random fiber grating ring," Opt. Lett., vol. 48, no. 9, p. 2389, May 2023.

[17] H. Pang et al., "Highly-stable optical injection wavelength locking of Er-doped random fiber lasers," J. Opt., vol. 25, no. 5, p. 055801, May 2023.

[18] H. Wu, B. Han, Z. Wang, and H. Liang, "Statistical properties of Er/Yb co-doped random Rayleigh feedback fiber laser," Chinese Opt. Lett., vol. 19, no. 2, p. 021402, 2021.

[19] J. Li et al., "Lévy spectral intensity statistics in a Raman random fiber laser," Opt. Lett., vol. 44, no. 11, p. 2799, Jun. 2019.

[20] G. P. Agrawal, Nonlinear fiber optics, 4th Editio. Academic Press, 2006.

[21] S. Lin et al., "Nonlinear dynamics of four-wave mixing, cascaded stimulated Raman scattering and self Q-switching in a common-cavity ytterbium/Raman random fiber laser," Opt. Laser Technol., vol. 134, p. 106613, Feb. 2021.

[22] A. S. L. Gomes, U. Österberg, and J. R. Taylor, "Spectral and temporal investigations of nonlinearities in a non-polarization preserving single-mode optical fibre," Appl. Phys. B Photophysics Laser Chem., vol. 41, no. 4, pp. 235–240, Dec. 1986.

[23] J. Bromage, "Raman Amplification for Fiber Communications Systems," J. Light. Technol., vol. 22, no. 1, pp. 79–93, Jan. 2004.

[24] Z. Wang et al., "Long-distance random fiber laser point sensing system incorporating active fiber," Opt. Express, vol. 24, no. 20, pp. 22448–22453, Oct. 2016.

[25] R Ma et al., "20 watt-level single transverse mode narrow linewidth and tunable random fiber laser at 1.5 μm band," Opt. Express vol. 30, no. 16, pp.28795-28804, Jun. 2022.